\documentclass{vgtc}                          

\ifpdf
  \pdfoutput=1\relax                   
  \usepackage{graphicx}                
  \DeclareGraphicsExtensions{.pdf,.png,.jpg,.jpeg} 
\else
  \usepackage{graphicx}                
  \DeclareGraphicsExtensions{.eps}     
\fi%

\graphicspath{{figures/}{pictures/}{images/}{./}} 

\usepackage{microtype}                 
\PassOptionsToPackage{warn}{textcomp}  
\usepackage{textcomp}                  
\usepackage{mathptmx}                  
\usepackage{times}                     
\usepackage{cite}                      
\usepackage{tabu}                      
\usepackage{booktabs}                  
\usepackage[normalem]{ulem}

\PassOptionsToPackage{hyphens}{url}\usepackage{hyperref}

\onlineid{0}

\vgtccategory{Research}

\vgtcinsertpkg

\title{LSDvis: Hallucinatory Data Visualisations in Real World Environments}

\author{Ari Kouts\thanks{e-mail: ari.kouts@viseo.com}\\ %
        \scriptsize VISEO %
\and Lonni Besan\c{c}on\thanks{e-mail: lonni.besancon@gmail.com}\\ %
     \scriptsize Link{\"o}ping University %
\and Michael Sedlmair\thanks{e-mail: Michael.Sedlmair@visus.uni-stuttgart.de}\\ %
     \scriptsize University of Stuttgart
\and Benjamin Lee\thanks{e-mail: Benjamin.Lee@visus.uni-stuttgart.de}\\ %
     \scriptsize University of Stuttgart}

\teaser{
  \centering
  \includegraphics[width=\linewidth]{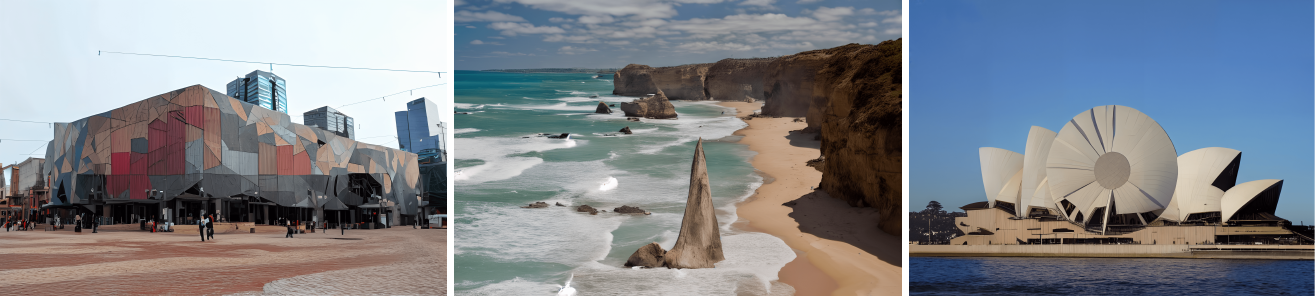}
  \caption{Examples of LSDvis based on Australian landmarks. Left: A bar chart of popular visiting times blended onto the facade of the Federation Square building. Middle: An area chart of visiting vehicle counts added as a rock in The Twelve Apostles. Right: A pie chart of revenue percentages blended into the shells of the Sydney Opera House.}
  \label{fig:teaser}
}

\abstract{We propose the concept of ``LSDvis'': the (highly exaggerated) visual blending of situated visualisations and the real-world environment to produce data representations that resemble hallucinations. Such hallucinatory visualisations incorporate elements of the physical environment, twisting and morphing their appearance such that they become part of the visualisation itself. We demonstrate LSDvis in a ``proof of proof of concept'', where we use Stable Diffusion to modify images of real environments with abstract data visualisations as input. We conclude by discussing considerations of LSDvis. We hope that our work promotes visualisation designs which deprioritise saliency in favour of quirkiness and ambience.
} 

\begin{document}


\firstsection{Introduction}

\maketitle

The conditions which data visualisations are viewed under are generally assumed (and subsequently ignored) by designers. High resolution display, well-lit room, the full attention of the reader, and so on. Of course, not all data is to be viewed in the comfort of one's own home or office. In numerous scenarios, it is advantageous to view data out in the real, physical places to which said data directly relates to, also known as \textit{situated visualisation} \cite{whiteSiteLensSituatedVisualization2009,willettEmbeddedDataRepresentations2017}. Situated visualisations are already ever-present in our day-to-day lives. Stationary 2D displays that show public transport arrival and departure times, waiting times for amusement park rides, and exchange rates at currency exchanges. These are all present-day examples of data being presented in situ to the general public.

With augmented reality (AR) headsets, researchers have recently begun investigating their use to support situated visualisation and analytics \cite{elsayedSituatedAnalytics2015,shinRealitySituationSurvey2023}. With AR, immersive visualisations can be directly situated or embedded onto or nearby the physical referents in the environment. This not only removes the need to rely on stationary displays, but also can reduce the level of spatial indirection between the data and the referent \cite{willettEmbeddedDataRepresentations2017}. However, existing research has arguably still kept to a very ``standard'' form of situated visualisation. A form that still uses many of the familiar idioms and techniques seen in so-called ``traditional'' data visualisation, such as the use of floating 2D panels with scatter plots and line charts (e.g.~ \cite{prouzeauCorsicanTwinAuthoring2020, fleckRagRugToolkitSituated2022}).

This approach is, of course, optimised for the viewability and understandability of the data in-situ. While perhaps an unassailable objective from a utilitarian standpoint, we argue that it does not make full use of the capabilities of AR to deliver highly engaging and embodied viewing of the data \cite{marriottImmersiveAnalytics2018}. In particular, one that leads to a stronger \textit{emotional response} \cite{wangEmotionalResponseValue2019, leeDataVisceralizationEnabling2021}.

So how can we elicit a stronger emotional response in people? The answer is simple: drugs\footnote{Disclaimer: We the authors do not condone the use of illicit drugs. We are just trying to be funny.}; specifically lysergic acid diethylamide or LSD. Ivan Sutherland \cite{sutherlandUltimateDisplay1965} back in 1965 had envisioned the \textit{Ultimate Display}: a computing system so powerful that it could control the existence of matter. A system that could surpass Metal Gear, breaking the laws of reality in order to present information and facilitate interactivity in ways not otherwise possible. While obviously not Sutherland's original intention, the manipulation and distortion of the real world to large extents can, if one stretches far enough, be seen as a form of hallucination.

So what happens when you combine hallucinations with data visualisation? This is where we propose the concept of \textit{LSDvis}: the (highly exaggerated) visual blending of situated visualisations and the real world to produce data representations that resemble hallucinations. Unlike regular situated visualisations, LSDvis is intended to be playful, letting people see data in ways which they have never seen before---especially when in the physical environments to which the data relates. Examples of which can be seen in Figure~\ref{fig:teaser}.

We illustrate the LSDvis concept by showcasing a gallery of AI-generated examples, which were created using a Stable Diffusion 1.5 model. Each example is based on an input 2D data visualisation and background image of some real world environment. We chose images in such a way to explore various conditions and determine in which scenarios our AI approach is successful. As this is a self-proclaimed ``proof of proof of concept'', we assume that such AI-generated images can simply be back-projected onto their real-world counterparts using AR, hence achieving a proper hallucinatory experience. Of course, further advancements in AI may soon allow these LSDvis images to be generated in real-time. We conclude by describing future steps for LSDvis, particularly in how it may play into its ``LSD'' moniker and psychedelics in general.

\section{Related Topics}
LSDvis is not an entirely new concept. From a visualisation perspective, we see it as an amalgamation of three distinct yet related topics: situated visualisation, data physicalisation, and ambient visualisation. We also discuss a fourth topic where others have presented data in non-conventional ways using AR and virtual reality (VR), and a fifth topic on generative models for visualisation which our work bears similarity to.

\subsection{Situated Visualisation}
As hallucinations are experienced by people in the real world, it makes sense to also view LSDvis in the context of the real world---hence, situated visualisation. Situated visualisation was first coined by White and Feiner~\cite{whiteSiteLensSituatedVisualization2009} to mean ``a visualisation that is related to and displayed in its environment.'' Willett et al.~\cite{willettEmbeddedDataRepresentations2017} later introduced physical referents, which are ``the real-world entities and spaces to which data corresponds.'' Since then, situated visualisation is now commonly associated with AR technologies \cite{bressaWhatSituationSituated2022}.

In existing work, situated visualisations serve to augment the physical referent through conventional visualisation idioms (e.g.~\cite{prouzeauCorsicanTwinAuthoring2020, fleckRagRugToolkitSituated2022, luoPearlPhysicalEnvironment2023}), or otherwise through abstract looking graphics (e.g.~\cite{skreinigARHeroGenerating2022,stanescuModelFreeAuthoringDemonstration2022}). The physical referent itself is visibly left unaltered. As LSDvis visually blends the data and the real world together however, the referent itself effectively becomes part (or the entirety) of the data visualisation---or at least it appears to be from the perspective of the viewer. This is the defining characteristic of LSDvis.


\subsection{Data Physicalisation}
As we are representing data with (parts of) real world objects, LSDvis is tangentially related to data physicalisation. Jansen et al.~\cite{jansenOpportunitiesChallengesData2015} defined data physicalisation as ``a physical artifact whose geometry or material properties encode data.'' The obvious difference of course is that data physicalisation is inherently tangible, thus allowing people to touch and feel data with their own bodies. Much like hallucinations are not actually tangible and exist only in one's perception, so too is LSDvis.
We don't see this intangibility as a strict disadvantage however. In some ways it expands our scope of possibilities even further, as physically morphing the real world to display data is, at least with current technology, impossible. Thus, what we can achieve with LSDvis is, at least in theory, limited to our own (sober) imaginations and the technology which we are using.

\subsection{Ambient Visualisation}
LSDvis also builds on the core ideals of ambient visualisation \cite{skogAestheticsUtilityDesigning2003}. Ambient visualisations are those which are visually integrated and blend in with the physical surroundings, such that people may be blissfully unaware of the visualisation's existence. Such visualisations may encode information that is based on the physical context, similar to situated visualisation. As an example, the \textit{Activity Wallpaper} by Skog~\cite{skogActivityWallpaperAmbient2004} visualises data based on local sound levels to indicate the level of physical activity in a room over a period of time. The visualisation simply looks like a typical wallpaper with stylistic patterns however, and not a traditional data visualisation with axes and labels.

This defocus on saliency is an intentional design goal, as people are not bombarded by hyper-salient, utility-driven information visualisations. Instead, they are treated to an aesthetically pleasing view of the data which they may have serendipitously noticed, thus sparking curiosity and emotional reactions \cite{rodgersExploringAmbientArtistic2011}. LSDvis aims to achieve a similar effect. While the distortion of the real world would inherently draw attention to the visualisation's existence, the visual blending may still cause the LSDvis to not be immediately noticeable, thus facilitating said serendipitous discovery.

\subsection{Alternative Data Representations in AR and VR}
Much like our work aims to represent data visualisations in a non-conventional form, so too has several works using AR or VR technologies---particularly by giving meaning to data through physical context. Lee et al.~\cite{leeDataVisceralizationEnabling2021} presented the notion of data visceralisation, which is the presentation of data in its original, physical form to enable the ``visceral understanding'' of data. Casamayou et al.~\cite{casamayouRideYourData2022} utilised a similar concept to represent temporal data as a roller coaster track which people can ride along in VR. In a similar vein, Assor et al.~\cite{assorExploringAugmentedReality2023} explored AR-based visualisations of waste data, such as by encoding the waste generated by a restaurant as virtual trash bags. In contrast, our work focuses on embedding data representations by directly altering the underlying image (and perhaps in the future, video or AR-stream sources). This is in some ways akin to large-scale projection mapping done on large buildings as public displays, but LSDvis instead morphs the structure of the environment rather than simply applying a decal over it \cite{leeDesignPatternsSituated2023}.

\subsection{Generative Models for Creative Visualisation}
In a visual essay at alt.VIS 2022, Wood~\cite{woodBeyondWalledGarden2022} discussed how AI can be used to make visualisations more expressive at the potential cost of their effectiveness. Our work takes on a similar mindset in that the ``walled garden'' of design rules and atomic variables need not be what we limit our creativities to---even if with the help of machines.

More broadly speaking, an increasingly large amount of work has been published on how models can be used to generate more creative visualisations. Many focus solely on the visual substitution of abstract glyphs and marks with more realistic or stylised objects or designs.
Zhang et al.~\cite{zhangDataQuiltExtractingVisual2020} showed how visual elements can be extracted from source images (e.g.~photos of coffee, sketches of juice boxes) to be then repurposed into custom pictorial visualisations.
Xiao et al.~\cite{xiaoLetChartSpark2023} also created pictorial visualisations, but instead by using a text-to-image generative model based on text inputs and the data of the inputted charts.
Wu et al.~\cite{wuViz2vizPromptdrivenStylized2023} took a similar approach by transforming an input 2D visualisation into a stylised form using textual prompts and a diffusion model.
Ying et al.~\cite{yingMetaGlyphAutomaticGeneration2023} instead automatically generate these glyphs through the use of metaphors that are derived from a spreadsheet.
Schetinger et al.~\cite{schetingerDoomDeliciousnessChallenges2023} identified a (non-exhaustive) list of potential usages of generative models for use in visualisation, to which many of the aforementioned works fall under the notions of design prettification or embellishment.
These types of work mainly seek to turn an existing 2D visualisation into something that is more visually stylised and appealing, usually by associating this style to the semantic meaning of the data (e.g.~data about glacier mass represented as a glacier \cite{xiaoLetChartSpark2023}). However, due to the situated nature of our work, we are constrained by the layout and design of real world environments \cite{leeDesignPatternsSituated2023}. Thus, the style is dictated more so by the physical environment itself and not of the underlying data semantics.

This notion of embedding data using the style and content of an existing, non-chart image can be seen in several other works. Tkachev et al.~\cite{tkachevMetaphoricalVisualizationMapping2022} use a visual style transfer on headshots of researchers to encode similarities in their research interests. The images of authors working on HCI might then be drawn in pencil style, while researchers working on computer graphics could be drawn in mosaic style. Coelho and Mueller~\cite{coelhoInfomagesEmbeddingData2020} introduced the idea of Infomages. Infomages refer to informal visual data representations that incorporate a data chart directly into a thematic image, as commonly done for InfoGraphics. A tool is provided that guides the user in creating such Infomages and that leverages object detection algorithms and style embedding techniques to avoid cumbersome manual image processing. The interactive tool supports the designer in carefully crafting InfoGraphics that merge subject matter images with the designer's interpretation of the data. In contrast to both these approaches, LSDvis takes a more general approach by blending data visualisations with any real world environment that may or may not be semantically related.

\section{Illustrating LSDvis}
We now go into detail illustrating our LSDvis concept. Given our targeted scope of this work, our goal was simply to create representative images of what we believe LSDvis may look like in the future. Thus, we turn to \sout{our machine overlords} an AI image generation model for our work, thereby exploring AI-based visualisation techniques similar to previous papers (e.g.~\cite{woodBeyondWalledGarden2022}). We first briefly describe our image generation process, then present the images which we have produced.

\subsection{AI Image Generation Process}
Each LSDvis was based on a 2D data visualisation and an image of a real world environment. We used a Stable Diffusion 1.5 model \cite{rombachHighResolutionImageSynthesis2022} with an img2img process. Two ControlNet conditional controls were added to each generation \cite{zhangAddingConditionalControl2023}. The first control was used to keep the style of the real world image intact. The second control was used to add the outline of the data visualisation, which could be one of the following: canny, hed, scribble, depth, softedge. These outlines were separate black and white images generated in a preprocessing stage that are machine readable by the ControlNet conditional controls.
Each generation was completed with a specific text prompt to create the output (e.g.~\textit{a detailed picture of a modern building with coloured bars on it}), which was manually reworded and refined over multiple trials to improve the image. Outputs which we were satisfied with were then upscaled using the SD upscale script (64 tiles, x4 upscale) using the R-ESRGAN\_4x+ upscaler model.

In the future, the prompt could instead be derived using a CLIP-like model \cite{radfordLearningTransferableVisual2021} or a multimodel LLM model \cite{openaiGPT4TechnicalReport2023}, creating a prompt based on what is in the two input images. This could also be combined with another LLM which can automatically recreate the prompt to integrate the data visualisation with the real world image in a certain way, thus removing the need for manual prompt creation.




\begin{figure*}[hbtp]
\centering
 \includegraphics[width=0.95\linewidth]{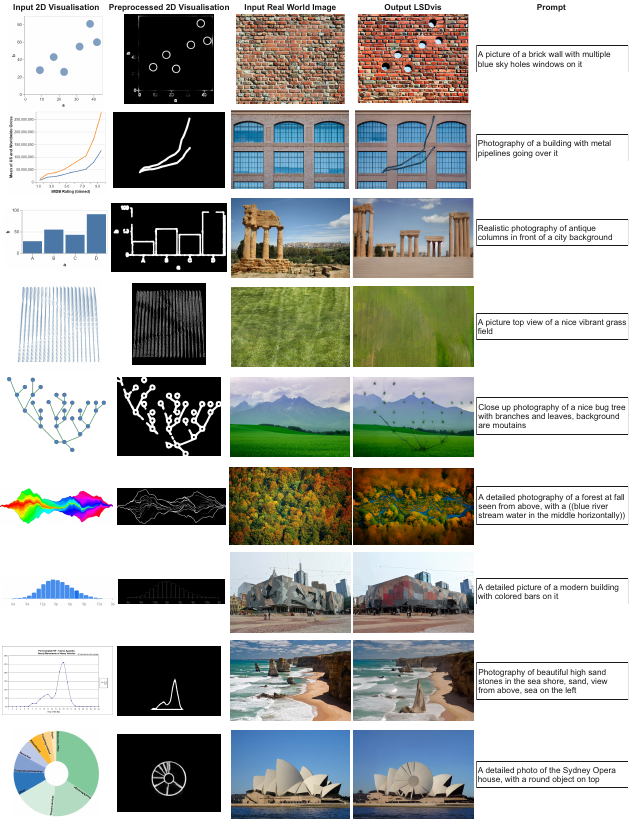}
\caption{Our example gallery of LSDvis. Each row represents a single LSDvis with its input 2D visualisation, the outline of the 2D visualisation after processing, its input real world image, and the output LSDvis that was generated based on the given prompt.}
\label{fig:big-figure}
\vspace{-5mm}
\end{figure*}

\subsection{Examples} \label{ssc:examples}
We now showcase several examples of LSDvis produced by our aforementioned AI generation process. We created these examples to test a variety of different conditions. First, we tested the effectiveness of additive and blending methods of image generation. Second, we experimented with literal representations of visualisation idioms. Lastly, we tested LSDvis with real world images we considered to be more difficult compared to the others, incorporating a level of situatedness to them.

We describe each example in the order which they appear in Figure~\ref{fig:big-figure}. This figure shows all of the inputs and the subsequent output for each example. We include footnotes to the sources of all images used in the generation where appropriate.

\subsubsection{Additive LSDvis}
\hspace{\parindent}\textbf{Scatter plot + Brick wall.} Being the first example we tested, this demonstrates how windows can be added to a brick wall\footnote{\url{https://www.patternpictures.com/red-brick-wall-2/}} in order to mimic a simple scatterplot\footnote{\label{veganote}Created using Vega-Lite \cite{satyanarayanVegaLiteGrammarInteractive2017}.}. Of course, while the number of data points is very small, we can imagine this to easily scale to larger datasets---simply by adding more windows! Note that this is the first of many examples which lose the detail of the axes and labels, which is likely one of the pitfalls of LSDvis.

\textbf{Line chart + Building facade.} Similar to the brick wall, this example adds metal pipes to a building's facade\footnote{\url{https://stocksnap.io/photo/building-facade-NJ7HAIDRGG}} based on a given line chart\textsuperscript{\ref{veganote}}. While the pipes themselves are not perfect, their resemblance is aided by the surrounding context that is the building. Of course, the colours of both lines are lost as a result of the generation, which is the first indication that LSDvis may not be suitable to encode colour. Regardless, we still see this as a success given the positional accuracy of the pipes.

\subsubsection{Blending LSDvis}
\hspace{\parindent}\textbf{Bar chart + Columns.} Compared to the previous two examples, this one more dramatically modifies the real world by blending and distorting many of its features. Thus, not only new columns\footnote{\url{https://www.wallpaperflare.com/brown-pillars-ruin-under-blue-skies-agrigento-sicily-archaeology-wallpaper-wynmj}} are added to the image based on a given bar chart\textsuperscript{\ref{veganote}}, but the existing column is modified to follow the new style. Such a dramatic blending of the real world may not be desirable in practice as it now becomes unrecognisable, but this is likely something which can be reined in with further optimisations to the prompt.

\textbf{Vector field + Grass.} This example blends a 2D vector field\textsuperscript{\ref{veganote}} with a top-down image of grass\footnote{\url{https://www.pxfuel.com/en/free-photo-oohea}}, in hopes of the output resembling a grass field that is being blown with the wind. As we can see, the output LSDvis does ever so slightly resemble the input vector field, but is sometimes difficult to make out due to the texture---particularly on the bottom right. 

\subsubsection{Literal LSDvis}
\hspace{\parindent}\textbf{Tree diagram + Field.} We experimented whether we could create an actual tree based on an input tree diagram\footnote{\url{https://stackoverflow.com/questions/8025342/undirected-graph-conversion-to-tree}} in an open field\footnote{\url{https://www.pxfuel.com/en/free-photo-jmtoz}}. While the resulting tree is rather skinny, it does resemble the input diagram remarkably well. LSDvis like this one are in a way similar to mirages, with physical (yet intangible) representations of data being added to the real world.

\textbf{Stream graph + Forest.} We also experimented with adding a stream graph\footnote{\url{https://stackoverflow.com/questions/13084998/streamgraphs-in-r}} as a literal stream inside of a forest\footnote{\url{https://www.pxfuel.com/en/free-photo-qnfrw}}. This would be much harder than the tree diagram, as the AI would need to not only add a river stream, but also blend it in with the existing forest to make it look natural. Needless to say it turned out very well. As mentioned above, the colour encoding does get lost as a result of this process. While the colour can potentially be added back in with further processing, seeing a stream with rainbow colours might be distressing to see, but arguably would add to the LSD effect.

\subsubsection{Situated LSDvis}
\hspace{\parindent}\textbf{Bar chart + Federation Square.} Federation Square\footnote{\url{https://commons.wikimedia.org/wiki/File:Federation\_Square\_\%28217189375\%29.jpeg}} is a popular landmark and venue for public events and festivals in Melbourne, Victoria, Australia. As it can get busy, knowing the number of visitors per hour is important, which can be represented as a bar chart\footnote{Image from Google Maps.}. While the AI does distort the shape of the building, it does a great job at making the bar chart appear as though it is actually part of it. This is, of course, partially aided by Federation Square having a rather abstract looking exterior, thus allowing the LSDvis to blend in much easier.

\textbf{Area chart + The Twelve Apostles.} This example is set in The Twelve Apostles\footnote{\url{https://commons.wikimedia.org/wiki/File:The\_twelve\_apostles\_Victoria\_Australia\_2010.jpg}} in Victoria, Australia. As it is in a remote area, visitors need to drive to see the natural landmark. Thus, this example shows the number of heavy vehicles which visit The Twelve Apostles, based on an input line chart\footnote{\url{https://www.researchgate.net/figure/Twelve-Apostles-Bus-Heavy-Vehicle-Hourly-Arrivals-and-Departures\_fig13\_283660861}}. This line chart is then converted into an area chart, represented as an additional rock formation in the LSDvis. As one can see, the result is impressive, although several rocks in the background were removed in the final image. With that all said, such an example may call into question the ethics of LSDvis (beyond its questionable name): does adding new elements to famous landmarks, akin to mirages, mislead the viewer as to its actual appearance? This is similar to the notion of \textit{visualisation mirages}~\cite{mcnuttSurfacingVisualizationMirages2020}, except causing (unintentional) deception of the real world and not of an inference made from the data. Perhaps intentionally having a less realistic LSDvis might avoid this issue entirely.

\textbf{Pie chart + Sydney Opera House}. The Sydney Opera House\footnote{\url{https://pixabay.com/photos/sydney-opera-house-australia-sydney-1223423/}} is arguably the most famous and recognisable Australian landmark. Other than being a work of art, it is also an actual business that hosts many shows such as, well, operas, and generates revenue. Therefore, this LSDvis incorporates a pie chart of the relative proportions of each of the Sydney Opera House's revenue streams\footnote{\url{https://www.sydneyoperahouse.com/about-us/how-we-work/governance-policies-and-corporate-information/annual-reports}}. The shells of the building are turned into the entire circle of the pie chart, with each segment roughly being visible. As with the other examples, the colour of the slices are lost in order to preserve the original colours and style of the building. Another potential question worth asking however is whether this distortion of famous landmarks can be seen as insensitive, especially if they have cultural or religious significance. This, however, we summarily decide not to answer.

\section{Discussion \& Future Work}
This work serves as merely a first step and ``proof of proof of concept'' to the concept of LSDvis. While it very much is intended to be lighthearted, we also believe that it demonstrates new possibilities and perspectives in how information could be presented to people. We first discuss several important topics that came about from reviews of this work, and then discuss the key considerations and future improvements to LSDvis.


\subsection{Leaning into the ``LSD'' moniker}

While the initial intention of the name LSDvis was to be catchy whilst referring to the blending of visualisations into the real world, it would seem that the overt reference to LSD elicited particularly insightful reviews and reflections on the concept that we now detail.

\textbf{Historical connotations of LSD.} Of particular interest, as reviewers had highlighted, the name LSDvis (incidentally) links back to a troubled history only slightly mitigated by the current positive research on the substance. Indeed, the history of LSD is a wild mixture of cultural appropriation of the use of psychoactive substances \cite{wadley2016psychoactive}, the devastating impact that the War on Drugs had on indigenous populations and minorities in the US \cite{provine2011race}, the unregulated and illegal research conducted by the CIA (e.g.~the MK Ultra project) \cite{mcwilliams1991covert}, and the unprecedented and mostly positively-reported impact the drug had on art production (e.g.~music) and culture \cite{wadley2016psychoactive}. Despite this, today LSD represents a promising drug for treating dependence \cite{krebs2012lysergic}, and displays interesting preliminary results on anxiety and depression \cite{holze2023lysergic}. Besides, LSD, like many other psychedelics, is considered to be physiologically safe if used within the standard doses. 
Pharmacologically speaking, LSD is considered to be non-addictive substance \cite{Halpern2010} and often ranked as one of the least harmful illegal drugs on the market, both for the users or society \cite{nutt2007development,nutt2010drug}. All in all, LSD may denote anything from a very troubled history to a promising or recreational experience. We however welcome the feelings that the naming of our concept evokes as it elicited interesting discussion points on the parallel one may do with the use of the drug.

\textbf{The hallucinatory metaphor.} The most common use of LSD ranges from micro-dosing to standard doses. The first usually elicits no real hallucinatory experience but may improve mood (or artistic) performance, while the second usually triggers a short-lived (six to twelve hours) hallucinatory experience, colloquially called ``trip''. Good trips may trigger short- to medium-lasting (weeks long) positive effects on users with respect to mood, sense of belonging, or spiritual enlightenment \cite{mcglothlin1967long}. In parallel, bad trips may evoke fears, anxiety, or a feeling of hopelessness. We believe that the response to LSD use may also characterise the wide range of responses an AR situated and blending visualization experience can evoke. From small and unrealistic incrustations of data representations that would clearly be identifiable as not belonging to reality (similar to micro-dosing), to particularly well-embedded representations that may evoke strong emotional responses on the user (similar to a trip), LSDvis may have a vastly different impact on users. Beyond even the characterisation of the experience on the ``trip spectrum'', one may also consider how good or bad the trip can be for users based on a multitude of parameters: realism, organicity of the visulisation, or even the data represented (e.g.~the data may generate climate anxiety \cite{clayton2020development}).

\textbf{Guides in LSDvis.} The hallucinatory experience is also interesting as it can link back to the idea that a guide may be necessary to maximise the experience. A guide is quite common to help users \textit{reflect} and \textit{cope} with the experience and limit the chances of a bad trip---a guide of LSDVis may have a very similar role. We imagine that the guide would be a visualisation-literate person or a domain expert of the visualised data who could answer the questions that the visualisation may elicit (reflect), provide more details on the dataset (reflect), assist the user with any negative feelings they may experience by ending the experience or providing more positive context and explanations (cope). In the case of generated climate anxiety for instance, one could point to positive data about climate change and its mitigation efforts. One may also consider that the guiding experience could be facilitated by an AI agent (based on LLMs for instance) that would manage to detect through simple biological measures (e.g.~heart-rate, eye movements) either: an unpleasant experience, and would then offer potential mitigation strategies (cope); or curiosity, which would then answer any user queries. Such an agent could even be made to blend with the hallucinatory experience of LSDvis.


\subsection{Considerations of LSDvis}
\hspace{\parindent}\textbf{Adding vs blending.} As mentioned in Section~\ref{ssc:examples}, our examples generally used one of two approaches: either adding new elements to the real world image, or distorting existing elements of the real world image. From our own observations, we feel that the additive method does yield more visually impressive results, at least given the method that we used. The blending method does result in more questionable outputs, over-distorting the real world image to the point where it may become unrecognisable. This was made apparent while creating our examples, as these blended images took much greater trial and error to achieve a reasonable output compared to additive images. While the additive method is not foolproof, as some examples had rocks removed or mountains reshaped, this can potentially be minimised via inpainting.

\textbf{Axes, labels, and colour.} Several of our examples had input data visualisations which included axes, labels, colour, text, etc. All of this information was lost during the image generation process, even if they were still present in the processed input. Obviously, such information can still be added with further processing, or in our case using Photoshop. However, it is worth questioning whether including these more abstract elements would ruin the aesthetics and appeal of LSDvis. For example, the multiple columns certainly do follow the same heights as the input bar chart, but without the $y$-axis it can be impossible to tell what data they encode (if any). From a utilitarian standpoint, this means LSDvis has no value, especially when viewed in the real world. From a more fun perspective, this means that LSDvis has an element of ``if you know you know''. Only those ``in the loop'' are aware of its meaning, which may make people more willing to engage with data and share their knowledge with others to flaunt this exclusive knowledge (similar to dank memes on the internet).

\textbf{Level of realism.} With the exception of the Sydney Opera House, all of our LSDvis examples do look fairly realistic even despite being AI-generated. While they may not be entirely practical (e.g.~randomly placed windows, irregularly bent pipes), they do still fall into the realm of plausibility. This therefore improves the ambience of LSDvis---the less bizarre it appears, the more it simply blends back into its surrounding environment. On the other hand, should the LSDvis appear to be entirely realistic (as is the case of the Sydney Opera House), it now enters the non-human uncanny valley whereby it very obviously encodes \textit{something} and might stand out almost too much. These two ends of the realism spectrum are in fact representations of hallucinatory experiences with LSD, from micro-dosing (using small doses) to trips. Therefore, if the goal of LSDvis is to subtly hide data in the real world without the viewer noticing, then aiming for a high level of realism is vital. If the goal is instead to just present data in a wacky and interesting manner, then go nuts---realism be damned.

\subsection{Improvements to LSDvis}
There are numerous obvious avenues for improvement to LSDvis. While the name and motivation of LSDvis may not be conducive towards more ``traditional'' research, all of these research directions are generalisable beyond our concept. Note that we intentionally do not include improvements that focus solely on the AI aspects.

\textbf{Taxonomisation of physical elements to visualisation components.} As was the original intention, many of our examples see physical elements being used to encode data. The same way that data can be mapped to graphics \cite{bertinSemiologyGraphics1983}, so to can data be mapped to physical elements (akin to data physicalisation \cite{jansenOpportunitiesChallengesData2015}). Of course, the scope of physical elements is orders of magnitude greater than simple graphical primitives, so we expect such a taxonomy to vary based on domain and use case.

\textbf{Real-time AR rendering.} While more of a technical limitation, the end goal would be to achieve such a hallucinatory effect in real-time AR. At present, each image takes approximately 20 to 30 seconds to generate, excluding any further iterations and refinement to optimise the prompt. Regardless, this is a real possibility given the current pace of AI research in the next several years.

\textbf{Animated transitions in LSDvis.} Similar to how animation can be a useful technique in data visualisation \cite{heerAnimatedTransitionsStatistical2007}, it may also be useful in LSDvis. Animations can further add to the hallucinatory effect of the LSDvis, as parts of the physical environment may move, pulsate, and twist. This may be creepy to some people, which depending on the use case might actually be the goal of LSDvis. Animation can also be used to visually transition the viewer between the real world and LSDvis, with them seeing the walls, furniture, and overall environment around them ``coming to life'' and give them the impression that they are, in a sense, falling down on spheres.

\section{Conclusion}
In this workshop paper we presented LSDvis, which is a lighthearted and playful way of visualising data in situated contexts in a manner that more closely resembles hallucinations. While perhaps lacking in its utility, we hope that LSDvis serves to promote visualisation designs which prioritise entertainment and engagement. In particular, we advocate for visualisations to not strongly pursue saliency and visual pop-out. Rather, visualisations---particularly those which are situated using AR---can instead be designed to be subtle, thus not bombarding the viewer with information on visually intrusive bright blue panels. This way, the information is there when the viewer notices it, but is otherwise ``invisible'', thus letting them focus on the real world.

\section{Conflict of Interest}
Lonni Besançon is a co-organiser of the 2023 alt.VIS workshop and was an organiser in 2021 and 2022.

\acknowledgments{
We would like to thank our reviewers for their insightful and thought-provoking feedback.
This project was funded by the German Research Foundation (DFG) project 495135767 and the Austria Science Fund (FWF) project I 5912-N (joint Weave project). The project is associated with and further supported by the DFG Excellence Cluster EXC 2120/1 – 390831618.
}

\bibliographystyle{abbrv-doi}

\bibliography{main}
\end{document}